\documentclass[fleqn]{article}
\begin{document}

\noindent
{\Large \bf Topological Charges, Prequarks and Presymmetry:  \\
a Topological Approach to Quark Fractional Charges}

\vspace{4mm}
\noindent
\emph{Ernesto A. MATUTE}
\vspace{4mm}

\noindent
\emph{Departamento de F\'{\i}sica, Universidad de Santiago de Chile,
Chile}

\noindent
E-mail: \emph{ematute@lauca.usach.cl}

\begin{abstract}
\noindent
A topological approach to quark fractional charges, based on charge
constraints unexplained by the Standard Model of particle physics, is
discussed.  Charge fractionalization is related to a tunneling process
occurring in time between pure gauge field configurations at the far past
and future associated with integer-charged bare quarks, named prequarks.
This transition conforms to a topologically nontrivial configuration of
the weak gauge fields in Euclidean space-time.  In this context, an
electroweak ${\cal Z}_{2}$ symmetry between bare quarks and leptons,
named presymmetry, is revealed.  It is shown that an effective
topological charge equal to the ratio between baryon number and the
number of fermion generations may be associated with baryonic matter.
The observed conservation of baryon number is then connected with the
conservation of this charge on quarks.  Similar results are
obtained for leptons in the dual scenario with local quark charges.
\end{abstract}

\section{Introduction}

At the level of the Standard Model of elementary particle physics,
matter consists of two kinds of particles: quarks and leptons,
which are divided into three generations.  They interact by means
of the strong, weak and electromagnetic forces.  These three
forces are described by theories of the same general kind: gauge
theories with local symmetries.  The gauge group is \linebreak
$SU(3)_{c} \times SU(2)_{L} \times U(1)_{Y}$, which provides with
the force particles: gluons, weak bosons and photons.
Table~\ref{Table1} gives a classification of the first generation
of quarks and leptons according to the representation they furnish
of this gauge symmetry group, where the $\nu_{eR}$ is introduced
because of the experimental signatures for neutrino masses, and
the conventional relation
\begin{equation}
Q = T_{3} + \frac{1}{2} \; Y
\label{charge}
\end{equation}
between electric charge and hypercharge is chosen.

\begin{table}[h]
\centering
  \caption{First generation of quarks and leptons.}   \label{Table1}
  \begin{tabular}{cccr}
  \hline \hline
  Fermions & $SU(3)_{c}$ & $SU(2)_{L}$ & $U(1)_{Y}$ \\
  \hline
  $\left( \begin{array}{cc}
          u_{L} \\ d_{L}
          \end{array}
  \right)$  & 3 & 2 & 1/3 \\
  $u_{R}$ & 3 & 1 & 4/3 \\
  $d_{R}$ & 3 & 1 & $-$2/3 \\
  $\left( \begin{array}{cc}
          \nu_{eL} \\ e_{L}
          \end{array}
  \right)$  & 1 & 2 & $-$1 \\
  $\nu_{eR}$ & 1 & 1 & 0 \\
  $e_{R}$ & 1 & 1 & $-$2 \\
  \hline \hline
  \end{tabular}
\end{table}

In the strong interaction sector there are substantial differences
between quarks and leptons: quarks come in triplets while leptons do
in singlets of the color group $SU(3)_{c}$.  However, in the electroweak
sector their properties are so similar that a deep connection between
them seems to exist.  Their appearance within each generation is
symmetrical: left-handed particles occur in doublets while right-handed
ones in singlets of $SU(2)_{L}$.  Their hypercharges are related: their
sum within each generation vanishes, i.e.,
\begin{equation}
3 \; [Y(u_{L}) + Y(d_{L}) + Y(u_{R}) + Y(d_{R})] +
Y(\nu_{eL}) + Y(e_{L}) + Y(\nu_{eR}) + Y(e_{R}) = 0 ,
\label{Ys}
\end{equation}
where the factor 3 takes into account the number of quark colors.  This
relation, magically obeyed, is crucial to cancel the triangle gauge
anomalies of the Standard Model.  Furthermore, it relates the number of
colors to the fractional nature of quark hypercharges.

Another striking pattern is the following one-to-one correspondence
between the quark and lepton hypercharges:
\begin{displaymath}
Y(u_{L}) = Y(\nu_{eL}) + \frac{4}{3} , \hspace{0.8cm} Y(u_{R}) =
Y(\nu_{eR}) + \frac{4}{3} ,
\end{displaymath}
\begin{equation}
Y(d_{L}) = Y(e_{L}) + \frac{4}{3} ,  \hspace{0.8cm} Y(d_{R}) =
Y(e_{R}) + \frac{4}{3} ,
\label{hyper}
\end{equation}
and similarly for the other two generations.  Quark and lepton
charges are similarly connected:
\begin{displaymath}
Q(u) = Q(\nu_{e}) + \frac{2}{3} , \hspace{0.8cm} Q(d) = Q(e) +
\frac{2}{3} .
\end{displaymath}
Equation (\ref{hyper}) is a remarkable and unexpected constraint, more
fundamental than (\ref{Ys}), which shows that the fractional hypercharge
of quarks relies just on a global 4/3 value, independent of flavor and
handness.  And, if the fractional hypercharge is related to the number
of colors as noted above, the factor 1/3 must be there because of quark
colors.  Besides, it clearly signals for a deeper discrete symmetry
between quarks and leptons.

The question is whether there is a way to understand these regularities
within the Standard Model.  The answer is yes, as I show in this talk,
just using the nontrivial topological properties of weak gauge field
configurations, as recently proposed~\cite{Matute:Matute}.  In
particular, it is seen that the factor 4 in (\ref{hyper}) is a
topological Pontryagin index fixed by the requirement of self-consistent
gauge anomaly cancellation, so that the number 4/3 is a universal value
which in a sense gives structure to the fractional hypercharge of quarks
and is conserved by the strong and electroweak interactions.

\section{Prequarks and presymmetry}

In order to understand the pattern in (\ref{hyper}) I pursue a
conventional procedure in field theory: start with integer bare local
charges for quarks and an exact electroweak symmetry between bare quarks
and leptons, and end with fractional charges and a broken or hidden
symmetry.  I refer to quarks with such bare charges as prequarks and
denote them by hats over the symbols that represent the corresponding
quarks.  The universal hypercharge shift 4/3 may be associated with a
bare structure that I denote by $X$.  Thus quarks may conveniently be
looked upon as ``made'' of the following bare mixtures:
\begin{equation}
u = \{ \hat{u} X \} , \hspace{0.8cm} d = \{ \hat{d} X \} ,
\label{composite}
\end{equation}
and similarly for the other two generations of quarks.  Neither new
physical fermions nor new binding forces underlying quarks are
introduced. I show that the structure $X$ is associated with an
Euclidean configuration of the standard gauge fields having topologically
nontrivial properties at infinity; in Minkowski space-time it represents
a tunneling process occurring in time by which prequarks get fractional
charges.  Leptons have integer charges with no topological contribution
from such a gauge field configuration.

What are the quantum numbers of $\hat{q}$ and $X$?  Each prequark has
the spin, isospin, color charge, and flavor of the corresponding quark.
Its bare weak hypercharge is the same as its lepton partner specified by
(\ref{hyper}) and its bare electric charge is defined according to
(\ref{charge}).  In this scenario all bare charges associated with local
fields are integral.  Finally, the baryon number of prequarks can be
fixed through the generalized Gell-Mann--Nishijima formula
\begin{equation}
Q = I_{z} + \frac{1}{2} (\mbox{B} + S + C + B^{*} + T) ,
\label{Nishi}
\end{equation}
where $I_{z}$, $S$, $C$, $B^{*}$, $T$ denote (pre)quark flavors.
Tables~\ref{Table2} and \ref{Table3} list the prequark quantum numbers.
Quarks and prequarks are identical in all their properties except
hypercharge, electric charge and baryon number assignments.

\begin{table}[h]
\centering
  \caption{First generation of prequarks.}    \label{Table2}
  \begin{tabular}{cccr}
  \hline \hline
  Fermions & $SU(3)_{c}$ & $SU(2)_{L}$ & $U(1)_{Y}$
  \\  \hline
  $\left( \begin{array}{cc}
          \hat{u}_{L} \\ \hat{d}_{L}
          \end{array}
  \right)$  & 3 & 2 & $-$1 \\
  $\hat{u}_{R}$ & 3 & 1 & 0 \\
  $\hat{d}_{R}$ & 3 & 1 & $-$2 \\
  \hline \hline
  \end{tabular}
\end{table}

\begin{table}[h]
\centering
  \caption{Prequark additive quantum numbers.}     \label{Table3}
  \begin{tabular}{lrrrrrr}
  \hline \hline
  Prequark & $\hat{u}$ & $\hat{d}$ & $\hat{s}$ & $\hat{c}$ &
  $\hat{b}$ & $\hat{t}$  \\  \hline
  B -- baryon number & $-$1 & $-$1 & $-$1 & $-$1 & $-$1 & $-$1 \\
  $Q$ -- electric charge & 0 & $-$1 & $-$1 & 0 & $-$1 & 0 \\
  $I_{z}$ -- isospin z-component & $\frac{1}{2}$ & $-\frac{1}{2}$ & 0 &
  0 & 0 & 0 \\
  $S$ -- strangeness & 0 & 0 & $-$1 & 0 & 0 & 0 \\
  $C$ -- charm & 0 & 0 & 0 & 1 & 0 & 0 \\
  $B^{*}$ -- bottomness & 0 & 0 & 0 & 0 & $-$1 & 0 \\
  $T$ -- topness & 0 & 0 & 0 & 0 & 0 & 1 \\
  \hline \hline
  \end{tabular}
\end{table}

Now regarding the bare quantum numbers associated with the
$X$-configuration, one gets, from~(\ref{hyper}), (\ref{charge}),
and (\ref{Nishi}),
\begin{displaymath}
Y(X) = \frac{4}{3} , \hspace{0.8cm}
Q(X) = \frac{1}{2} Y(X) = \frac{2}{3} , \hspace{0.8cm}
\mbox{B}(X) = 2 Q(X) = \frac{4}{3} .
\label{Xcharges}
\end{displaymath}
When these fractional charges are added to the prequark bare ones, the
fractional quark charges are obtained.

On the other hand, as readily seen from Tables~\ref{Table1} and
\ref{Table2}, a discrete ${\cal Z}_{2}$ symmetry between prequarks and
leptons is disclosed in the electroweak sector of the Standard Model.
Specifically, this symmetry, which I refer to as presymmetry, means
invariance of the classical electroweak Lagrangian of prequarks and
leptons under the transformation
\begin{displaymath}
\hat{u}^{i}_{L} \leftrightarrow \nu_{eL} , \hspace{0.8cm}
\hat{u}^{i}_{R} \leftrightarrow \nu_{eR} , \hspace{0.8cm}
\hat{d}^{i}_{L} \leftrightarrow e_{L} , \hspace{0.8cm}
\hat{d}^{i}_{R} \leftrightarrow e_{R} ,
\end{displaymath}
and similarly for the other generations, where $i$ denotes the color
degree of freedom.  Observe that prequarks and leptons have the same
$\mbox{B}-\mbox{L} = -1$; i.e., $\mbox{B}-\mbox{L}$ is the right fermion
number to be considered under presymmetry.  Besides, it is important to
realize that presymmetry is not an ad hoc symmetry; as (\ref{hyper})
shows, it underlies the electroweak relationships between quarks and
leptons.  Also, prequarks do not define a new layer of the structure of
matter; as shown in the following section, they are just bare quarks
whose charges are normalized by the universal contribution of a
topologically nontrivial Chern--Simons configuration of the standard
gauge fields.

\section{Topological currents and gauge anomaly cancellation}

Following the standard work~\cite{Matute:Weinberg}, it is found that the
$U(1)_{Y}$ gauge current for prequarks and leptons
\begin{displaymath}
\hat{J}^{\mu}_{Y} = \overline{\hat{q}}_{L} \gamma^{\mu} \frac{Y}{2}
\hat{q}_{L} + \overline{\hat{q}}_{R} \gamma^{\mu} \frac{Y}{2}
\hat{q}_{R} + \overline{\ell}_{L} \gamma^{\mu} \frac{Y}{2} \ell_{L} +
\overline{\ell}_{R} \gamma^{\mu} \frac{Y}{2} \ell_{R} ,
\end{displaymath}
with $\hat{q}_{L}(\ell_{L})$ and $\hat{q}_{R}(\ell_{R})$ uniting
the left-handed and right-handed prequarks (leptons) in all
generations, respectively, exhibits the $U(1)_{Y} \times
[SU(2)_{L}]^{2}$ and $[U(1)_{Y}]^{3}$ gauge anomalies generated by
the integer hypercharge of prequarks:
\begin{equation}
\partial_{\mu} \hat{J}^{\mu}_{Y} = - \frac{g^{2}}{32 \pi^{2}} \;
\left[ \sum_{\hat{q}_{L} , \ell_{L}} \frac{Y}{2} \right] \;
\mbox{tr} \; W_{\mu\nu} \tilde{W}^{\mu\nu} - \frac{{g'}^{2}}{48 \pi^{2}}
\; \left[ \sum_{\hat{q}_{L} , \ell_{L}} \biggl( \frac{Y}{2} \biggr)^{3} -
\sum_{\hat{q}_{R} , \ell_{R}} \biggl( \frac{Y}{2} \biggr)^{3} \right] \;
F_{\mu\nu} \tilde{F}^{\mu\nu} ,
\label{anomaly1}
\end{equation}
where $g$, $g'$ and $W_{\mu\nu}$, $F_{\mu\nu}$ are the $SU(2)_{L}$,
$U(1)_{Y}$ couplings and field strengths, respectively.  The anomalies
are introduced because the sums in (\ref{anomaly1}) do not vanish:
\begin{displaymath}
\sum_{\hat{q}_{L}, \ell_{L}} Y = - 8 \; N_{g} , \hspace{0.8cm}
\sum_{\hat{q}_{L} , \ell_{L}} Y^{3} - \sum_{\hat{q}_{R} , \ell_{R}} Y^{3}
= 24 \; N_{g} ,
\end{displaymath}
where $N_{g}$ is the number of fermion generations.

Now, as often noted in the literature~\cite{Matute:Weinberg}, the terms
on the right-hand side of (\ref{anomaly1}) are divergences of
gauge-dependent currents:
\begin{equation}
\partial_{\mu} \hat{J}^{\mu}_{Y} = - \frac{1}{2} \;
\left( \sum_{\hat{q}_{L} , \ell_{L}} \frac{Y}{2} \right) \;
\partial_{\mu} K^{\mu}
- \frac{1}{2} \; \left[ \sum_{\hat{q}_{L} , \ell_{L}} \biggl(
\frac{Y}{2} \biggr)^{3} - \sum_{\hat{q}_{L} , \ell_{L}} \biggl(
\frac{Y}{2} \biggr)^{3} \right] \; \partial_{\mu} L^{\mu} ,
\label{div}
\end{equation}
where
\begin{equation}
K^{\mu} = \frac{g^{2}}{8 \pi^{2}} \epsilon^{\mu\nu\lambda\rho} \;
\mbox{tr} \; \left( W_{\nu} \partial_{\lambda} W_{\rho} - \frac{2}{3} i g
W_{\nu} W_{\lambda} W{\rho} \right) , \hspace{0.8cm}
L^{\mu} = \frac{{g'}^{2}}{12 \pi^{2}} \epsilon^{\mu\nu\lambda\rho}
A_{\nu} \partial_{\lambda} A_{\rho} ,
\label{CS}
\end{equation}
which are the Chern--Simons classes or topological currents related to
the $SU(2)_{L}$ and $U(1)_{Y}$ gauge groups, respectively.
Equation~(\ref{div}) can be rewritten in the form
\begin{displaymath}
\partial_{\mu} \hat{J}^{\mu}_{Y} = - N_{\hat{q}} \; \partial_{\mu}
J^{\mu}_{X} ,
\end{displaymath}
where $N_{\hat{q}} = 12 \, N_{g}$ is the number of prequarks and
\begin{equation}
J^{\mu}_{X} = \frac{1}{4 N_{\hat{q}}} K^{\mu}
\sum_{\hat{q}_{L} , \ell_{L}} Y + \frac{1}{16 N_{\hat{q}}} L^{\mu}
\left( \sum_{\hat{q}_{L} , \ell_{L}} Y^{3} -
\sum_{\hat{q}_{R} , \ell_{R}} Y^{3} \right)
= - \frac{1}{6} \, K^{\mu} + \frac{1}{8} \, L^{\mu}
\label{currX}
\end{equation}
is the current to be associated with the $X$-configuration of gauge
fields introduced in (\ref{composite}).  It can be combined with the
anomalous fermionic current $\hat{J}^{\mu}_{Y}$ to define a new current
\begin{equation}
J^{\mu}_{Y} = \hat{J}^{\mu}_{Y} + N_{\hat{q}} \; J^{\mu}_{X} ,
\label{newJ}
\end{equation}
which is conserved but it is gauge dependent.

The local counterterms to be added to the Lagrangian of prequarks and
leptons, needed to obtain the anomaly-free current of (\ref{newJ}), are
given by
\begin{equation}
\Delta {\cal L} = g' N_{\hat{q}} \, J^{\mu}_{X} A_{\mu} ,
\label{Lagran}
\end{equation}
so that, due to the antisymmetry of $\epsilon^{\mu\nu\lambda\rho}$, only
the non-Abelian fields will be topologically relevant, as expected.  They
produce nontrivial effects as described in the following.

The charge corresponding to the current in (\ref{newJ}) is
\begin{equation}
Q_{Y}(t) = \int d^{3}x \; \hat{J}^{o}_{Y} + N_{\hat{q}} \int d^{3}x \;
J^{o}_{X} ,
\label{Qt}
\end{equation}
which is not conserved after all because of the existence of the
topological charge associated with gauge fields.  To see this, the change
in $Q_{Y}$ between $t = - \infty$ and $t = + \infty$ is calculated.  As
usual, it is assumed that the region of space-time where the energy
density is nonzero is bounded~\cite{Matute:Jackiw}.  Therefore this
region can be surrounded by a 3-dimensional surface on which the field
configuration becomes pure gauge, i.e.,
\begin{displaymath}
W_{\mu} = - \frac{i}{g} (\partial_{\mu} U) U^{-1} .
\end{displaymath}
This field can be obtained from $W_{\mu}=0$ by a transformation $U$ that
takes values in the corresponding gauge group.  In this case, using
(\ref{Qt}), (\ref{currX}) and (\ref{CS}) for the gauge fields, one ends
up with
\begin{displaymath}
Q_{Y}(t) = \frac{N_{\hat{q}}}{6} \; n_{W}(t) ,
\end{displaymath}
where
\begin{equation}
n_{W}(t) = \frac{1}{24 \pi^2} \int d^{3}x \; \epsilon^{ijk} \;
\mbox{tr} \; (\partial_{i}U U^{-1} \partial_{j}U U^{-1}
\partial_{k}U U^{-1})
\label{winding}
\end{equation}
is the winding number of the non-Abelian gauge transformation.
This number is integer-valued if we consider a fixed time $t$ and
assume that \(U(t,\mbox{\boldmath $x$})\) equals a direction
independent constant at spatial infinity, e.g., $U \rightarrow 1$
for \(|\mbox{\boldmath $x$}| \rightarrow \infty\). The usual
argument to see this property is based on the observation that
this $U$ may be viewed as a map from the 3-dimensional space with
all points at infinity regarded as the same onto the 3-dimensional
sphere of parameters $S^{3}$ of the $SU(2)_{L}$ group manifold.
But 3-space with all points at infinity being in fact one point is
topologically equivalent to a sphere $S^{3}$ in Minkowski space.
Therefore $U$ determines a map $S^{3} \rightarrow S^{3}$.  These
maps are characterized by an integer topological index which
labels the homotopy class of the map.  This integer is
analytically given by (\ref{winding}).  For the Abelian case,
$n_{W}=0$.

Thus, for the field configurations joined to prequarks which at the
initial $t = - \infty$ and the final $t = + \infty$ are supposed of the
above pure gauge form, the change in charge becomes
\begin{equation}
\Delta Q_{Y} = Q_{Y}(t=+\infty) - Q_{Y}(t=-\infty) =
\frac{N_{\hat{q}}}{6} \left[ n_{W}(t=+\infty) - n_{W}(t=-\infty)
\right] .
\label{deltaQ1}
\end{equation}
The difference between the integral winding numbers of the pure
gauge configurations characterizing the gauge fields at the far
past and future can be rewritten as the topological charge or
Pontryagin index defined in Minkowski space-time by
\begin{equation}
Q_{T} = \int d^{4}x \; \partial_{\mu} K^{\mu} = \frac{g^{2}}{16
\pi^{2}} \int d^{4}x \; \mbox{tr} \; (W_{\mu\nu}
\tilde{W}^{\mu\nu}) , \label{QT1}
\end{equation}
where it is assumed that $K^{i}$ decreases rapidly enough at spatial
infinity.  This topological index is gauge invariant, conserved and,
for arbitrary fields, can take any real value.  But, as shown above,
for a pure gauge configuration it is integer-valued.  Thus
\begin{equation}
Q_{T} = n = n_{W}(t=+\infty) - n_{W}(t=-\infty) ,        \label{QT2}
\end{equation}
so one has from (\ref{deltaQ1}) that
\begin{equation}
\Delta Q_{Y} = N_{\hat{q}} \; \frac{n}{6} .              \label{deltaQ2}
\end{equation}
These equations have a special significance when the integral in
(\ref{QT1}) is analytically continued to Euclidean space-time. The
integral topological charge remains the same but it can now be
associated with an Euclidean field configuration that at this
point we identify with the $X$ in (\ref{composite}).  In this
case, the meaning of (\ref{QT2}) is that the Euclidean
$X$-configuration interpolates in imaginary time between the real
time pure-gauge configurations in the far past and future which
are topologically inequivalent.  The condition is like the one
established for instantons.  Thus the continuous interpolation is
to be considered as a tunneling process, so that a barrier must
separate the initial and final gauge field configurations. It
should also be noted that if such an analytical continuation to
Euclidean space-time is ignored, nonzero topological charge then
implies nonvanishing energy density at intermediate real time and
so no conservation of energy.

A consequence of (\ref{deltaQ2}) is that $N_{\hat{q}}$ prequarks have
to change their $Y/2$ in the same amount $n/6$.  For each prequark it
implies the hypercharge change
\begin{equation}
Y_{\hat{q}} \rightarrow Y_{\hat{q}}  + \frac{n}{3} .     \label{norma}
\end{equation}
Therefore, the nontrivial topological properties of the $X$-configuration
give an extra contribution to prequark local hypercharge, so that the
integer bare values one starts with have to be shifted.  Accordingly, the
gauge anomalies have to be re-evaluated.  It is found that anomalies are
cancelled self-consistently for $n=4$, the number of prequark flavors per
generation, since now
\begin{equation}
\sum_{q_{L}, \ell_{L}} Y = 0 , \hspace{0.8cm} \sum_{q_{L} ,
\ell_{L}} Y^{3} - \sum_{q_{R} , \ell_{R}} Y^{3} = 0 .
\label{sumsY}
\end{equation}

The above hypercharge normalization with topological charge $n=4$
is consistent with (\ref{hyper}) and it means restoration of gauge
invariance, breaking of the electroweak presymmetry in the Abelian
sector, dressing of prequarks into quarks, and the substitution of
the bare presymmetric model by the Standard Model. In fact, from
(\ref{sumsY}) and (\ref{currX}) one notes that the gauge-dependent
topological current $J^{\mu}_{X}$ associated with the
$X$-configuration is cancelled.  However, the corresponding
conserved topological charge is gauge independent and manifests
itself as a universal part of the prequark hypercharge according
to (\ref{hyper}).  More precisely, if the topological current
introduced in (\ref{Lagran}) and its induced hypercharge obtained
in (\ref{deltaQ2}) are considered, an effective prequark current
$\hat{J}^{\mu}_{Y,\mbox{eff}}$ can be defined by
\begin{displaymath}
\hat{J}^{\mu}_{Y,\mbox{eff}} = \frac{2}{3} \;
(\overline{\hat{q}}_{L} \gamma^{\mu} \hat{q}_{L} +
\overline{\hat{q}}_{R} \gamma^{\mu} \hat{q}_{R})
\end{displaymath}
to absorb the nontrivial effects of such $N_{\hat{q}} J^{\mu}_{X}$
current, namely, cancellation of gauge anomalies and inducement of the
hypercharge $4/3$ on prequarks regardless of flavor and handness.  Upon
using this, (\ref{newJ}) becomes
\begin{displaymath}
J^{\mu}_{Y} = \overline{\hat{q}}_{L} \gamma^{\mu} \frac{Y+4/3}{2}
\hat{q}_{L} + \overline{\hat{q}}_{R} \gamma^{\mu} \frac{Y+4/3}{2}
\hat{q}_{R} + \overline{\ell}_{L} \gamma^{\mu} \frac{Y}{2} \ell_{L} +
\overline{\ell}_{R} \gamma^{\mu} \frac{Y}{2} \ell_{R} .
\end{displaymath}
At this point, prequarks with fractional hypercharge, which includes the
universal $4/3$ part, have to be identified with quarks.  The replacement
of prequarks by quarks in the strong, weak and Yukawa sectors is
straightforward as they have the same color, flavor and weak isospin.

All of this is essentially done at the level of the classical
Lagrangian where the specific field configuration $X$, which only
mixes with prequarks, is used.  Again, this configuration is a
pseudoparticle in Euclidean space-time and a tunneling process in
Minkowski space-time by which a prequark hypercharge changes from
integer to fractional values. It is also interesting to note that
as stated by the model self-consistency is the reason for the
``magical'' cancellation of gauge anomalies in the Standard Model.
Moreover, the factor 1/3 in (\ref{norma}) is due to the number of
prequark colors (assuming same number of prequark and lepton
families) introduced in (\ref{currX}) through $N_{\hat{q}}$, which
predicts, as expected, that quarks carry 1/3-integral charge
because they have three colors.

As it is shown above, standard bare quarks instead of prequarks
are the fermions to start with in the quantum field theory
treatment. The novel news is that bare quarks have fractional
charges owing to a universal contribution from a classical gauge
field configuration with specific topological properties (i.e.,
$n=4$). Underlying this charge structure one has presymmetry as
reflected in (\ref{hyper}); an electroweak ${\cal Z}_{2}$ symmetry
between integer-charged bare quarks and leptons which is broken by
the vacuum configuration of gauge fields but it accounts for the
electroweak similarities between quarks and leptons.

\section{Effective topological charge and baryon number}

The underlying topologically nontrivial gauge field configuration
in bare quarks suggests to associate an effective topological
charge with baryonic matter.  If it is considered that an
effective fractional topological charge equal to
$Q_{T}=n/N_{\hat{q}}=1/N_{c}N_{g}=1/9$, where $N_{c}$ is the
number of colors, can be associated with each $X$-configuration,
then, within the bare configuration of (\ref{composite}), it is
natural to define an effective topological charge for quarks
which is conserved and should be related to its fermion number. As
a general rule it is found that
\begin{equation}
Q_{T} = \frac{\mbox{B}}{N_{g}} .
\label{rule}
\end{equation}
Thus $\Delta$B = $N_{g} \Delta Q_{T}$ for any baryon-number violating
process, i.e., only topological effects may violate baryon number.

To see the consistency of the definition for the charge in
(\ref{rule}), the baryon plus lepton number (B+L) violating
processes induced non-perturbatively by electroweak instanton
effects may be considered.  According to
Ref.~\cite{Matute:tHooft}, for three generations, one electroweak
instanton characterized by a topological charge $Q_{T}=1$ is
associated with quark and lepton number violations in three units:
$\Delta$B=$\Delta$L=$-$3, matching three baryons or nine quarks
and three antileptons.  A decay such as $p+n+n \rightarrow
\mu^{+}+\bar{\nu}_{e}+\bar{\nu}_{\tau}$ is then allowed.  Within
the bare configuration scheme of (\ref{composite}), this rule
implies that one instanton induces a process in which nine
$X$-configurations vanish.  The nine $X$'s make precisely the same
topological charge of the electroweak instanton.  In a sense, the
definition in (\ref{rule}) brings back topological charge
conservation in quantum flavor dynamics.  In Minkowski space-time
one electroweak instanton corresponds to a process which has
associated the topological charge change $\Delta Q_{T} = 1$.  It
induces a process with the effective topological charge change
$\Delta Q_{T} = -1$ associated with the vanishing of nine quarks
and baryon number violation $\Delta \mbox{B} = -3$.  It should
also be noted that consistency between the topological charge of
the instanton and the effective one assigned to quarks
corroborates the above value $n=4$ of the Pontryagin index fixed
by gauge anomaly constraints.

Finally, it can be seen from (\ref{rule}) that a baryon number
violation  $\Delta$B=$-$1 in actual experiments means an effective
topological-charge change $\Delta Q_{T} = - 1 / N_{g}$.  In particular,
proton decay would imply the presence of a background gauge source with
topological charge $1 / N_{g}$~\cite{Matute:Farhi}.  Stability of a free
proton is expected anyway because instanton-like events cannot change
topological charge by this amount.  For three generations, the effective
topological charge has a value $Q_{T}=1/9$ for quarks and $Q_{T}=1/3$
for nucleons.

\section{Conclusion}

Insights into the classical dynamics of the standard weak gauge
fields which give to quarks fractional charge relative to leptons
have been given.  A self-consistent method to adjust bare local
charges with Chern--Simons contributions has been pursued.
Presymmetry, a discrete ${\cal Z}_{2}$ symmetry between bare
quarks and leptons, has been introduced to understand quark-lepton
similarities in the electroweak sector of the Standard Model.  I
have also presented arguments, based on the topological character
of quark fractional charges and conservation of the associated
effective topological charge, to explain the observed conservation
of baryon number and predict stability of a free proton.

In this talk I have emphasized a picture in which the integral
hypercharges of leptons are entirely associated with local fields.
An alternative, dual point of view is to consider quark fractional
hypercharges as the local ones, so that the integral hypercharge of
leptons relies, as seen from (\ref{hyper}), just on a universal $-$4/3
value which also has a topological character.  In such a case, this
number 3 by which the Pontryagin index $n=4$ is divided cannot
correspond to the number of colors because color is not a property of
leptons.  The number of generations is the only available degree of
freedom.  This interpretation solves the fermion family problem:
presymmetry requires that the number of fermion generations be equal to
the number of quark colors.  On the other hand, following an analysis
similar to the one presented for baryons, it is concluded  that leptonic
matter has associated an effective topological charge given by $Q_{T} = -
\mbox{L} / N_{g}$.  Thus, the general result appears to be
\begin{displaymath}
Q_{T} = \frac{\mbox{B}-\mbox{L}}{N_{g}} .
\end{displaymath}
This charge is absolutely conserved, so that $\Delta
(\mbox{B}-\mbox{L}) = 0$ in any physical process, as expected.

\subsection*{Acknowledgments}

This work was supported in part by the Departamento de Investigaciones
Cient\'{\i}ficas y Tecnol\'o-
\linebreak                  gicas, Universidad de Santiago de Chile.

\end{document}